\documentclass[11pt,a4paper]{article}
\usepackage[utf8]{inputenc}
\usepackage[T1]{fontenc}
\usepackage{lmodern}
\usepackage{microtype}
\usepackage[margin=2.5cm]{geometry}
\usepackage{setspace}
\usepackage{parskip}
\usepackage{graphicx}
\usepackage{booktabs}
\usepackage{enumitem}
\usepackage{hyperref}
\hypersetup{colorlinks=true,linkcolor=blue,citecolor=blue,urlcolor=blue}
\usepackage{titling}
\usepackage{fancyhdr}
\usepackage{caption}
\usepackage{amsmath,amssymb}
\usepackage{natbib}
\usepackage{multicol}
\usepackage{aas_macros}

\pretitle{\vspace*{-1cm}\begin{center}\LARGE\bfseries}
\posttitle{\par\end{center}\vskip 0.5em}
\preauthor{\begin{center}\large}
\postauthor{\end{center}}
\predate{\begin{center}\large}
\postdate{\end{center}}
\fancypagestyle{titlepageheader}{
  \fancyhf{} 
  \fancyhead[L]{ESO White Paper Call: Expanding horizons}
  
}

\renewcommand{\footnoterule}{
  \kern -3pt
  \hrule width \textwidth
  \kern 2.6pt
}

\begin{document}


\begin{titlepage}
  \thispagestyle{titlepageheader}
  {\LARGE \textbf{Exploiting tidal asteroseismology in binary populations from combined space photometry and time-resolved high-resolution spectroscopy}\par}
  
  \vspace{0.5cm}

  \textbf{Authors:} Ema Šipková\footnote{Corresponding author: ema.sipkova@kuleuven.be} (KU Leuven, Belgium), Alex Kemp (KU Leuven, Belgium), Dario Fritzewski (KU Leuven, Belgium), Andrew Tkachenko (KU Leuven, Belgium), Dominic M. Bowman (Newcastle University, UK), Conny Aerts (KU Leuven, Belgium), Jasmine Vrancken (KU Leuven, Belgium)

  \textbf{Keywords:} Asteroseismology -- Stars: oscillations -- Binaries: close -- Methods: observational -- Techniques: spectroscopic
  
  \vfill

  \begin{figure}[h!]
      \centering
      \includegraphics[width=0.8\linewidth]{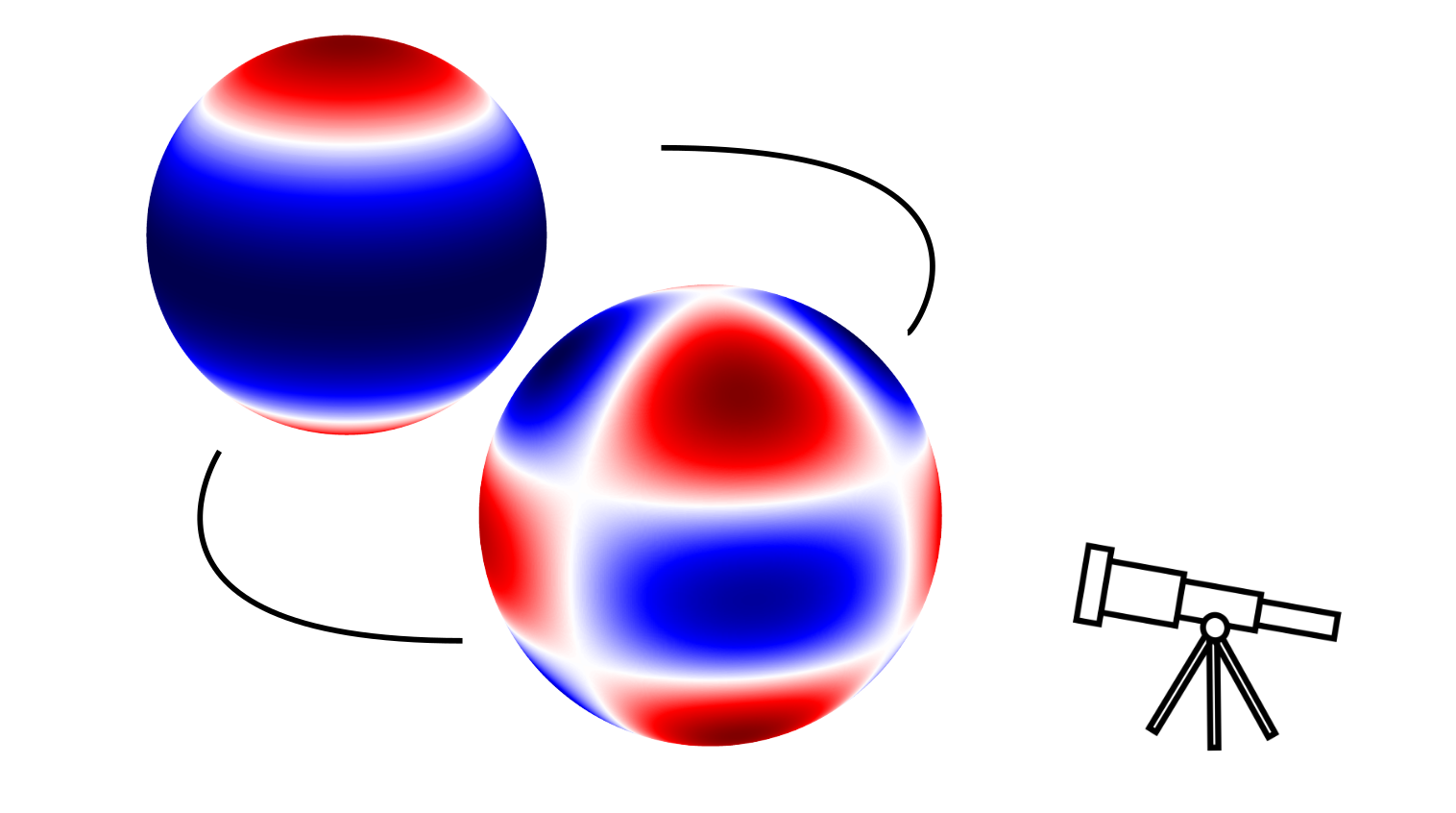}
      \caption*{\centering \small Illustrative schematic of a close binary system with pulsating stellar components (pulsation modes: $l=2,\,m=0$ and $l=4,\,m=2$). Credit: J. Vandersnickt}
  \end{figure}
  
  \vfill

  \centering
  \textbf{Abstract}
  
  \begin{minipage}{0.9\textwidth}
    \small
    Space-based photometry has substantially increased the number of pulsating stars found in binary systems by more than four orders of magnitude. Combined with high-resolution spectroscopy, high-precision photometry offers model-independent constraints on stellar parameters and internal processes. The advent of space-based photometric surveys has given us access to populations of tidally perturbed pulsators, which offer a unique and demanding set of constraints on tidal physics and stellar interiors. However, we lack the ability to undertake multi-epoch, high-resolution spectroscopy at large scale. The ability to obtain phase-resolved, high-resolution spectra would allow us to place precise, model-independent constraints on the stellar properties of pulsators in binary systems that will truly test our close binary asteroseismic modelling techniques, leading to much-needed constraints on fundamental stellar and binary physics. The need to properly cover the large parameter-space of binary stars demands a large-scale, population-level analysis in order to understand the complex landscape of binary stellar evolution. To enable this population-level analysis, we need a dedicated multi-fibre spectrograph (30–200 fibres) with high spectral resolution ($R\geq50000$), high signal-to-noise ratio ($\mathrm{S/N\geq300}$), and a limiting magnitude of approximately 15. Such a spectrograph would be capable of efficiently resolving the pulsation variability on the order of minutes and orbit motion on the order of days to years for many targets.
  \end{minipage}
  \vspace{0.1cm}

\end{titlepage}


\section{Introduction and Background}
\label{sec:intro}
\subsection{Studying binarity and pulsations}
Binary systems\footnote{In this paper, we only describe binary systems, but our instrument proposal also applies to higher-order multiples (triples, quadruples, etc.); they only require long-term monitoring} contain two stars orbiting a common center of mass. Space-based photometry provides continuous, high-precision observations of the brightness variations of millions of star systems, resulting in the detection and characterization of a large number of eclipsing binaries (EBs). Eclipses offer strong geometric constraints on the system geometry, including the orbital inclination, relative radii, and luminosity ratios. Orbital-phase resolved, low/medium-resolution spectroscopy provides complementary radial velocity (RV) measurements that determine orbital motion and mass ratios, while high-resolution spectroscopy is needed for abundances and captures line-profile variations (LPVs) that trace surface structure, rotation, and the pulsation geometry of numerous oscillation modes.

Frequencies of pulsation modes depend on internal structure, composition gradients, rotation, and magnetic fields. In this way, stellar pulsations can serve as probes of internal structure. This is the basis of asteroseismology, but its success heavily depends on the ability to identify the pulsation modes' geometries.
Pressure (p) modes propagate both in convective and radiative regions with the pressure gradient acting as the dominant restoring force, whereas gravity (g) modes are predominantly restored by buoyancy and only propagate in radiative regions.
In this paper, we consider the pulsating components to be early-type stars (stars with spectral classes O, B, A and F). These also show higher multiplicity fractions than late-type stars \citep[e.g.][]{duchene2013}.
In addition to photometric variability, non-radial pulsations generate LPVs, which provide direct information on the pulsation geometry and mode identification. With the benefit of mode identification from LPVs, asteroseismic modelling can provide extremely precise constraints on masses, radii, ages, internal rotation, and mixing processes with high precision \citep[e.g.,][]{aerts2021}. These constraints can then be confronted with the model-independent constraints from binarity to perform extremely rigorous tests of models of binary evolution.

Combining the information provided by eclipses, RVs, and pulsations allows for a comprehensive, tightly constrained description of both the global stellar parameters and the details of the internal structure. This synergy significantly reduces modelling uncertainties and degeneracies. It enables tests of stellar structure and evolution theory, as well as tidal theory with unprecedented accuracy.
A reliable estimate of the fraction of pulsating stars in binary systems requires an observationally complete mapping of a broad parameter space, covering different pulsator classes, orbital configurations, and stellar evolutionary stages. This will improve our models and allow us to characterise and robustly constrain the interplay between pulsations and binarity on a population level.

\subsection{Effect of binarity on pulsations}
A companion star in a binary system exerts a gravitational force that distorts its structure, elongating it along the line connecting the two centers of mass. The strength of tidal interaction depends primarily on the ratio of the stellar radius to the orbital separation \citep{zahn1977,Hut1981}. Tides facilitate the exchange of angular momentum between the star's spin and orbital motion, driving spin-orbit alignment and synchronisation, as well as the circularisation of eccentric binaries. This response is highly dependent on the star’s internal structure and it significantly alters its structure and evolution.

Tides can also lead to a variety of phenomena that influence pulsations in a star \citep[see reviews by][]{southworth-bowman2025, aertstkachenko2024}. \citet{kumar1995} theorised that stellar deformation driven by the dynamical tide in eccentric binaries can act as a driving mechanism for tidally excited oscillations (TEOs) at frequencies that are exact multiples of the orbital period. Space-based photometric surveys have provided extensive observational evidence for this mechanism through the discovery of eccentric systems that exhibit characteristic periastron brightening accompanied by oscillation frequencies that coincide exactly with multiples of orbital frequency \citep{thomspon2012}.

In close binary systems, tides can also affect pulsations in a variety of ways. Tidally perturbed pulsators can exhibit several distinct phenomena. These include tidal tilting, where the pulsation axis is displaced toward the orbital axis; tidal trapping, where mode visibility is confined to one hemisphere of the star; and tidal amplification, where pulsation amplitudes are preferentially enhanced on the side facing the companion \citep[see][]{fuller2021}. These effects can observationally manifest as frequency splitting linked to the orbital geometry, the presence of multiple effective pulsation axes, or amplitude and phase modulation during the orbit. Because the specific signatures are sensitive to mode geometry, internal structure, and the strength of tidal deformation, such systems provide valuable constraints on internal structure and rotation, while simultaneously introducing complexities that require advanced modelling approaches to fully exploit these effects.

\subsection{Modelling techniques}
State-of-the-art modelling techniques currently exist separately for EBs, such as WD \citep{wd}, ELLC \citep{ellc}, and PHOEBE 2.2+ \citep{prsa2016}, and for stellar pulsations, e.g. FAMIAS \citep{zima2008}. Modelling the light curves of pulsating EBs therefore proceeds in two steps: (a) construct a binary model, and b) subtract this model from the observations and apply pulsation-modelling techniques on residual data. This approach is effective only when pulsation amplitudes are small relative to eclipse depths and when pulsation and orbital frequencies are sufficiently separated. In systems where pulsations dominate the light curve, the procedure can be reversed, with pulsations modelled first and binary effects fitted to the residuals. Systems with equal contribution from eclipses and pulsations are probably among the hardest to model \citep[see ][]{southworth-bowman2025}.

A more robust approach would involve a unified framework that simultaneously models the binary orbit and the pulsations, fitting light curves, radial velocities, and pulsation-induced line-profile variations in a self-consistent model. The next step in modelling could also include spectral synthesis of binary components and comprehensive description of spectral time-series data, similar to the light curve.

By the 2040s, self-consistent model of pulsations in binary systems needs to be developed to provide further insight into the complexity of the problem and advance the modelling. The comprehensive analysis of pulsating EB systems using the developed models will require high-resolution, short-cadence spectroscopy capable of resolving variability in individual spectral lines. Such spectroscopic data need to provide the cadence and signal-to-noise required to characterise LPVs, separate pulsation and orbital contributions and study the effects of binarity on pulsation modes.

\section{Open Science Questions in the 2040s}
\label{sec:openquestions}
Pulsating binaries provide valuable constraints on stellar structure and evolution by enabling independent measurements of fundamental parameters and reducing uncertainties in asteroseismic modelling. The interaction between pulsations and binarity is, however, inherently complex. Addressing this complexity in the coming decades requires progress on several key issues:
\begin{itemize}
    \item Expand the number of pulsating stars in binary systems by several orders of magnitude and characterise how different classes of pulsators behave under different binary configurations.
    \item Begin observational mapping of tidal effects across a broad parameter space, covering a range of mass ratios, orbits, evolutionary states, and pulsation types.
    \item Quantify how tides modify mode geometry and visibility, including the occurrence of tidally excited and tidally perturbed pulsations.
    \item Investigate the effect of spin–orbit coupling on pulsation modes when alignment, differential rotation, and synchronisation timescales are included.
    \item Develop robust methods to disentangle pulsations from binarity by combining LPVs, RV measurements, and high-precision space-based photometry in one global consistent framework.
\end{itemize}

\section{Technology and Data Handling Requirements}
\label{sec:tech}
A comprehensive study of pulsations in binary systems requires instrumentation capable of obtaining high-resolution, high signal-to-noise spectroscopy for large samples (tens of thousands) of targets and over the full range of relevant timescales. Resolving pulsation variability demands cadences on the order of minutes, while resolving binary orbits requires coverage from days to years. To answer all questions in Section~\ref{sec:openquestions}, the sample of pulsating binaries will require analysis of tens of thousands systems to cover the pulsational and orbital parameter space. Current spectroscopic facilities do not provide the necessary combination of resolution ($R \geq 50 000$), cadence, and long-term stability, to provide high S/N observations ($\mathrm{S/N\geq300}$) for such a large number of targets. To provide complementary data to a large number of targets observed with space telescopes, the limiting magnitude of the spectrograph should be at least $\sim$15~mag. A multi-object spectrograph (30–200 fibres) that delivers all of these capabilities in a single instrument is therefore essential for population-level studies of tidal interactions and pulsation behaviour across a broad range of binary configurations.

A Technology Development Team (TDT) needs to define these technical specifications, develop the spectrograph concept, design robust and scalable analysis pipelines, and establish archiving standards that ensure long-term accessibility and interoperability of the resulting data products, making full use of modern machine-learning contexts.

\section*{References}
{
\begin{multicols}{2}
\scriptsize
\begingroup
\twocolumn
\renewcommand{\section}[2]{}%
\bibliographystyle{aa}
\bibliography{references}
\endgroup
\end{multicols}
}

\end{document}